\documentclass[twocolumn]{aastex631}

\defcitealias{OwenEstrada}{OC20}

\shorttitle{Modeling Radius Valley Emergence}
\shortauthors{VanWyngarden \& Cloutier}

\begin{document}

\title{Modeling Multiple Radius Valley Emergence Mechanisms With Multi-Transiting Systems}

\correspondingauthor{Madison\ VanWyngarden}
\email{madiv@bu.edu}

\author[0000-0002-7456-2167]{Madison VanWyngarden}
\affiliation{Boston University, 233 Bay State Rd, Boston, MA 02215, USA}
\affiliation{Center for Astrophysics $|$ Harvard \& Smithsonian, 60 Garden St, Cambridge, MA 02138, USA}
\author[0000-0001-5383-9393]{Ryan Cloutier}
\affiliation{Department of Physics \& Astronomy, McMaster University, 1280 Main St W, Hamilton, ON, L8S 4L8, Canada}
\affiliation{Center for Astrophysics $|$ Harvard \& Smithsonian, 60 Garden St, Cambridge, MA 02138, USA}

\begin{abstract}
Close-in planets smaller than Neptune form two distinct populations composed of rocky super-Earths and sub-Neptunes that may host primordial H/He envelopes. The origin of the radius valley separating these two planet populations remains an open question and has been posited to emerge either directly from the planet formation process or via subsequent atmospheric escape. Multi-transiting systems that span the radius valley are known to be useful diagnostics of XUV-driven mass loss. Here, we extend this framework to test XUV-driven photoevaporation, core-powered mass loss, and an accretion-limited primordial radius valley model. Focusing on multi-transiting systems allows us to eliminate unobservable quantities that are shared within individual systems such as stellar XUV luminosity histories and the properties of the protoplanetary disk. We test each proposed radius valley emergence mechanism on all 221 known multi-transiting systems and calculate the minimum masses of the systems' enveloped planets to be consistent with the models. We compare our model predictions to 75 systems with measured masses and find that the majority of systems can be explained by any of the three proposed mechanisms. We also examine model consistency as a function of stellar mass and stellar metallicity but find no significant trends. More multi-transiting systems with mass characterizations are required before multi-transiting systems can serve as a viable diagnostic of radius valley emergence models. Our software for the model evaluations presented herein is available on GitHub and may be applied to future multi-transiting system discoveries.

\end{abstract}


\section{Introduction} \label{sec:intro}
The discovery of thousands of exoplanets with NASA’s Kepler space telescope has revolutionized modern understanding of exoplanet demographics. One key finding is that planets with radii between 1 and 4 $R_{\oplus}$ represent the most common planets around FGKM dwarfs within $\sim 100$-day orbital periods \citep[e.g.][]{Youdin2011, Howard2012, DressingCharbonneau, Fressin2013, Petigura2013, MortonSwift}. A closer look at the occurrence of sub-Neptune sized planets, following improved stellar characterization, revealed the existence of a \emph{radius valley}: a bimodal distribution of planetary radii with a dearth of planets between $\sim 1.6-1.9 R_{\oplus}$ \citep{Fulton2017,FultonPetigura2018,VanEylen2018,CloutierMenou,Hardegree2020}. The subpopulation of super-Earths below the radius valley appear to have terrestrial compositions \citep{Weiss2014,Rogers2015,Dressing2017,LuquePalle}, whereas the larger sub-Neptunes exhibit lower bulk densities that may be explained by terrestrial cores that are enveloped in H/He envelopes with mass fractions of $\sim1\%$ \citep{OwenWu,Gupta2019,Wu2019,Bean2021,Lee2022}. 

The origin of the radius valley remains an open question. Prominent theories of the origin of the radius valley can be broadly classified as either arising directly from the planet formation process or from atmospheric escape. In the former class of models, super-Earths and sub-Neptunes form as separate populations, which may proceed from terrestrial planet formation in a gas-depleted environment \citep{LopezRice}, dry versus water-rich formation and migration \citep[e.g.][]{Raymond2018,Venturini2020}, or from limited gas accretion onto low mass cores that gives rise to a primordial radius valley \citep{LeeConnors, Lee2022}. Conversely, atmospheric escape models assume that all small planets accrete gaseous envelopes that can be subsequently stripped by a thermally-driven mass loss mechanism, leaving behind a bare, rocky core. This atmospheric escape may occur during the first $\sim 100$ Myrs around Sun-like stars by XUV-driven photoevaporation \citep{OwenJackson, OwenWu}, or it may be driven by core-powered mass loss wherein the cooling luminosity of a planetary core after formation drives atmospheric escape over Gyr timescales \citep{Gupta2019, Gupta2020, Ginzburg2018}. 

Many recent studies have sought to distinguish between the variety of proposed radius valley emergence mechanisms. Examples include searches for population-level trends as functions of planet radius, instellation, stellar mass, and age  \citep{Berger2020, Loyd2020, Rogers2021, Petigura2022, Berger23, VanEylen2021, Cherubim_2023,Bonfanti2023}. Other examples include modeling different regimes in which XUV-driven escape dominates over CPML, or vice-versa \citep{OwenSchlichting}, searching for signatures of atmospheric escape \citep{GuptaSchlichting}, and comparative studies of multi-transiting systems (\cite{OwenEstrada}; hereafter \citetalias{OwenEstrada}). Systems of multiple transiting planets that span the radius valley provide unique tests of radius valley emergence mechanisms because they can be used to marginalize over stellar and protoplanetary disk parameters that are common between planets in the same system. Many of these marginalized parameters are not directly observable, which makes evaluating physical models dependent on parameter assumptions. For example, the photoevaporation mechanism depends on the XUV luminosity of the host star, particularly within the first $\sim 0.1-1$ Gyrs, depending on the host star's mass \citep{OwenJackson, ShkolnikBarman14, ShneiderShkolnik}. The vast majority of known planetary systems are older than these timescales such that their host stars' XUV luminosities $L_{XUV}$ at the early stages relevant for XUV-driven photoevaporation are inaccessible. Assessing the importance of photoevaporative evolution in individual planetary systems is therefore sensitive to assumptions on $L_{\mathrm{XUV}}$. 
The unique opportunity afforded by multi-transiting systems in mitigating these uncertainties was first argued by \citetalias{OwenEstrada}, who developed the \texttt{EvapMass} numerical model to test the photoevaporation mechanism using well-characterized multi-transiting systems.

In this study, we build upon previous work by extending the numerical \texttt{EvapMass} model to consider a larger suite of radius valley emergence models that includes XUV-driven photoevaporation, core-powered mass loss, and the primordial radius valley model. Our paper is structured as follows: in Section~\ref{sec: data} we discuss each of the aforementioned models, Section~\ref{sec:code} presents our publicly-available code to evaluate each model, while Section~\ref{sec:methods} discusses our selection of multi-transiting systems and the application of our framework to those systems. We discuss our findings in Section~\ref{sec:discussion} and conclude with a summary in Section~\ref{sec:conclusion}. The code used to generate each figure is available on \href{https://mvanwyngarden.github.io/PEPPER_figures}{GitHub}. 

\section{Radius Valley Emergence Models} \label{sec: data}
Here we detail our adopted assumptions regarding planetary structure, the planet's thermal evolution, and the three radius valley emergence mechanisms that we consider in this study.

\subsection{Planetary Structure} \label{sec: planetstructure}
Two types of planetary bulk compositions are considered in this study: terrestrial bulk compositions (deemed ``rocky'') and H/He-enveloped terrestrials (deemed ``enveloped''). 
We calculate rocky core masses using interior structure models for a fully differentiated, two-layered, solid planet with a fixed iron core-mass fraction of 33\% \citep{Zeng2013}. Our assumed value of 33\% is supported by the empirical compositional link between stellar refractory abundances and rocky exoplanet core-mass fractions that is seen in the solar system and in many exoplanetary systems \citep{Adibekyan2021}. This is expected because major refractories that make up super-Earth interiors (Mg, Si, Fe) condense out of the nebular gas at similar temperatures \citep{Lodders2003} and the relative abundances of these materials is remarkably uniform among stars in the solar neighbourhood \citep{Bedell2018}. For enveloped planets, the core is topped with a two-layered, solar composition envelope structure\citep[see][]{OwenWu}, which consists of a deep adiabatic layer and an upper, low-mass isothermal layer at the planet's equilibrium temperature. The two layers are separated at a radiative-convective boundary (RCB), whose radius is $R_{\mathrm{RCB}}$. The more massive adiabatic layer is approximated as containing the entire mass of the envelope. 

\subsection{Thermal Evolution} \label{sec: thermalevol}
Primordial atmospheres will undergo thermal contraction over time. This implies that planets' H/He envelopes were larger at early times. This effect is significant when comparing two planets within the same system in order to marginalize over unknown parameters. Following \citetalias{OwenEstrada}, we must select a specific epoch at which to compare the two planets to account for the contracting radii of the radiative-convective boundaries and ensure that we are comparing planets at the same stage in their thermal evolution. However unlike \citetalias{OwenEstrada}, we must a select different timescale for each emergence model that corresponds to the relevant physics of the model.    
For XUV-driven photoevaporation, we choose this timescale to be the saturation timescale of the host star ($t_{\mathrm{sat}}$). While $t_{\mathrm{sat}}$ is a function of stellar mass, there is no well-defined monotonic trend between these values \citep{McDonald2019}. We therefore assign fixed values of $t_{\mathrm{sat}}$ in three stellar mass bins. For planets orbiting stars of $M_\star>0.8$, a time of 100 Myrs is used, for planets around stars with $M_\star<0.3$, 1 Gyr is used, and for planets orbiting stars of an intermediate-mass a time of 300 Myrs is used \citep{McDonald2019}. For core-powered mass loss, we choose a time of 1 Gyr, which is the approximate time over which this mechanism is likely operating \citep{Gupta2019}. Lastly for the primordial radius valley mechanism, we select a fiducial value of 1 Myr. This value is chosen from models of cooling-limited gas accretion in \cite{Lee2019} which predict that the accretion rate onto low mass cores falls off past 1 Myr.   

\subsection{XUV-driven Photoevaporation} \label{section:PE}
Here we consider XUV-driven photoevaporation (i.e. PE) as a process that has been posited to explain the emergence of the radius valley by stripping the primordial envelopes of gas-enveloped terrestrial cores, leaving behind a bare rocky super-Earth. Photoevaporation is driven by XUV heating by the host star, which can drive a hydrodynamic outflow on orbiting planets \citep{Jackson2012, OwenWu13, OwenWu, Jin2014, LopezFortney,ChenRogers, JinMordasini}.  The photoevaporation component of our model's framework follows largely from the \texttt{EvapMass} code \citetalias{OwenEstrada}. We summarize our model below and highlight any differences between \texttt{EvapMass} and our implementation. 

For photoevaporation to plausibly explain the architecture of any multi-transiting rocky-enveloped planet pair, the photoevaporative mass loss timescale of the system's gas-enveloped planet must exceed that of the system's rocky planet (i.e. $t_{\dot{M},\mathrm{env}}/t_{\dot{M},\mathrm{rocky}}> 1$). As a result, equating the two timescales (i.e. $t_{\dot{M},\mathrm{env}}=t_{\dot{M},\mathrm{rocky}}$ allows us to calculate the minimum enveloped planet mass required for photoevaporation to explain the system architecture. The timescale over which a planet loses its atmosphere due to photoevaporation is

\begin{equation}
{t_{\dot{M},\mathrm{PE}}} \propto \frac{a^2\, M_p^2\, X_{\mathrm{env}}}{\eta\, R_c^3\, L_{\mathrm{XUV}}},
\label{eq:tmlPE}
\end{equation}

\noindent where $a$, $M_p$, $X_{\mathrm{env}}=M_{\mathrm{env}}/M_{\mathrm{core}}$, and $R_c$ are the planet's semimajor axis, mass, envelope mass fraction at the time of formation, and core radius, respectively. The parameters $\eta$ and $L_{\mathrm{XUV}}$ represent the efficiency of atmospheric mass loss and XUV luminosity of the host star, respectively. We highlight that Eq.~\ref{eq:tmlPE} need only to be written as a proportionality because we are only concerned with the ratio of mass loss timescales between planets in multi-transiting systems. Furthermore, $L_{\mathrm{XUV}}$ is treated as a constant in Eq.~\ref{eq:tmlPE} because it is a property of the host star that is therefore common between the system's planets.

During preliminary tests of our model, we considered three different prescriptions for the mass loss efficiency $\eta$. These prescriptions included 1) a constant value, 2) the following scaling with escape velocity: $\eta \propto v_{\mathrm{esc}}^{-2}$ \citep{OwenWu}, and 3) an interpolation of hydrodynamic mass loss rate calculations from \cite{OwenJackson}. The first prescription has been commonly adapted in the literature (e.g. \citealt{OwenWu}, \citetalias{OwenEstrada}), but when applied to the diversity of known multi-transiting systems, unphysical values of $\eta > 1$ are commonly produced. Consequently, 
we performed a set of preliminary tests to compare the results of our model when adopting a constant efficiency parameter of $\eta =0.1$ versus the mass loss rates from hydrodynamical simulations \citep{Jackson2012}. Our findings revealed that our model predictions are only marginally sensitive to this choice of efficiency parameter. For example, we found that differences in the enveloped planets' minimum masses exhibited a median level of variation at only 1.7\% across all known multi-transiting planets. Consequently, we report the results from the constant efficiency parameter.

\subsection{Core-Powered Mass Loss} \label{section:CPML}
The second mechanism that we consider in this study is core-powered mass loss (i.e. CPML), wherein the luminosity of a newly formed planetary core can drive thermal atmospheric escape over Gyr timescales \citep{Gupta2019, Gupta2020, Ginzburg2018}. Following disk dispersal, the outer layers of the planet's atmosphere can become hydrodynamically unstable against escape as energy from the cooling core diffuses across the RCB, heating the upper layers of the atmosphere. Similar to photoevaporation, core-powered mass loss is a thermally-driven mass loss mechanism, so we can evaluate the consistency of multi-transiting systems under CPML just as we did for XUV-driven escape in Section ~\ref{section:PE}. Our model builds upon the simplified, analytical model from \cite{Cloutier-Eastman}. 

The mass loss rate under core-pore powered mass loss is limited by the cooling luminosity of the planet and the escape rate of molecules at the Bondi radius, the radial distance at which the escape velocity is equal to the local sound speed. The mass loss timescale can be expressed as

\begin{equation}
t_{\dot{M},\mathrm{CPML}}=\frac{M_{\mathrm{env}}}{\mathrm{min}(\dot{M}_{\mathrm{env}}^E, \dot{M}_{\mathrm{env}}^B)\eta},
\label{eq:tmlCPML}
\end{equation}

\noindent where the envelope mass $M_{\mathrm{env}}$ is divided by the minimum of either the cooling-energy limited mass loss rate, $\dot{M}_{\mathrm{env}}^E$, or the mass loss rate at the Bondi radius, ${\dot{M}_{\mathrm{env}}^B}$. The mass loss efficiency is not well-constrained for this mechanism, so we adopt the same strategy as in our photoevaporation calculations. Namely, a constant efficiency value.

The energy-limited mass loss rate $\dot{M}_{\mathrm{env}}^E$ is constrained by the thermal and gravitational energy available for cooling, given by

\begin{equation}
E_{\mathrm{cool}} = g\Delta R\left(\frac{\gamma}{2\gamma-1}M_{\mathrm{env}} + \frac{\gamma-1}{\gamma(\gamma_c -1)}\frac{\mu}{\mu_c}M_{\mathrm{core}}\right),
\end{equation}

\noindent where $\gamma$ is the atmospheric adiabatic index, and $\gamma_c$ is the core adiabatic index, we take $\gamma=7/5$ and $\gamma_c=4/3$, following \citep{GuptaSchlichting}. To determine whether the envelope will survive, the cooling energy can be compared to the energy required to lose the atmosphere: 

\begin{equation}
E_{\mathrm{loss}} = \frac{GM_{\mathrm{core}}M_{\mathrm{env}}}{R_c}.
\end{equation}

\noindent The loss of pressure support from the dispersing protoplanetary disk causes the planetary envelope to shrink on the timescale of disk dispersal until its thickness $\Delta R = R_p-R_c$ is approximately equal to the core radius \citep{Ginzburg2016, OwenWu}. Once the envelope radius is equal in size to the core radius, it is either lost or maintained based on the envelope mass fraction $X_{\mathrm{env}}$. As demonstrated in \cite{Ginzburg2018}, if $X_{\mathrm{env}}$ exceeds a threshold value of $\simeq 5\%$, then $E_{\mathrm{cool}} < E_{\mathrm{loss}}$ and the system will not have enough energy to continually lose mass. However, if $X_{\mathrm{env}}$ is less than this threshold value, then $E_{\mathrm{cool}} > E_{\mathrm{loss}}$ and atmospheric mass loss will occur in a run-away process, rapidly evolving the planet into a rocky super-Earth. The cooling-energy limited mass loss rate is given by

\begin{equation}
\dot{M}_{\mathrm{env}}^E=\frac{L\eta}{gR_c} \label{eq:MdotE}
\end{equation}

\begin{equation}
L=-\frac{dE_{\mathrm{cool}}}{dt}=\frac{64\pi}{3}\frac{\sigma T_{\mathrm{RCB}}^4R_B}{\kappa\rho_{\mathrm{RCB}}} 
\end{equation}

\noindent where $T_{\mathrm{RCB}} \sim T_{\mathrm{eq}}$ is the temperature at the RCB because the radiative layer is isothermal. $R_B$ is the modified Bondi radius, $\kappa$ is the opacity at the RCB, and $\rho_\mathrm{RCB}$ is the density at the RCB \citep{Ginzburg2016}.  Atmospheric mass loss is also limited by the escape rate of molecules at the speed of sound at the Bondi radius. The mass loss rate at the Bondi radius is

\begin{equation}
\dot{M}_{\mathrm{env}}^B=4\pi R_s^2c_s\rho_{\mathrm{RCB}}\mathrm{exp}{\left(\frac{-GM_p}{c_s^2R_{\mathrm{RCB}}}\right)},
\label{eq:MdotB}
\end{equation}

\noindent where $R_s= GM_p/2c_s^2$ is the sonic point and $c_s$ is the local iosthermal sound speed, $c_{s}=\sqrt{k_bT_{\mathrm{eq}}/ \mu}$ at a fixed atmospheric mean molecular weight $\mu=2.35$ (i.e. solar composition).

The minimum of the two rates given by Eqs.~\ref{eq:MdotE} and~\ref{eq:MdotB} is used to compute the mass loss timescale. However, for the systems considered in this work, we find that the escape rate at the Bondi radius is always the minimum of the two mass loss rates such that we rewrite Eq.~\ref{eq:tmlCPML} as 

\begin{equation}
t_{\dot{M},\mathrm{CPML}}=\frac{M_p}{\dot{M}_{\mathrm{env}}^B\eta}(t).
\end{equation} 

\noindent The enveloped planet's minimum core mass to be consistent with core-powered mass loss is the value of $M_p$ for which the mass loss timescales of the system's rocky and enveloped planets are equal.

\subsection{Primordial Radius Valley}
The third and final radius valley emergence mechanism that we consider is the model of a primordial radius valley \citep{LeeConnors, Lee2022}. This mechanism is unique from the two prior models as it considers how the radius valley could have arisen directly from the planet formation process without relying on atmospheric escape to transform gas-enveloped sub-Neptunes into rocky super-Earths. Contrary to the underlying assumption of photoevaporation and core-powered mass loss, the primordial radius valley mechanism does not assume that all small close-in planets initially formed as enveloped terrestrials. Instead, super-Earths and sub-Neptunes form simultaneously and are distinguished by their differing amounts of accreted gas, where accretion is limited by the protoplanet's core mass and other thermodynamic properties of the disk, as discussed below.

For a primordial radius valley to emerge, the gas mass that is accreted onto a solid protoplanet is set by the maximally cooled isothermal mass $M_{\mathrm{iso}}$ \citep{LeeChiang}. This is the envelope mass at which the entire envelope thermally relaxes with the surrounding nebular material and becomes isothermal at the local disk temperature. At this stage, the core is unable to cool further and the planet can no longer accrete additional nebular material. The dependence of the isothermal mass on factors such as the planet's core mass and the local disk temperature $T_{\mathrm{disk}}$ is what limits gas accretion and produces the population of gas-poor super-Earths without the need to invoke any form of atmospheric mass loss. The maximally cooled isothermal mass is given by

\begin{equation}
M_{\mathrm{iso}} \propto \rho_{\mathrm{disk}} \int_{R_{\mathrm{core}}}^{R_{\mathrm{out}}} r^2\mathrm{exp}\left[\frac{M_{\mathrm{core}}}{c^2_{s}}\left(\frac{1}{r}-\frac{1}{R_{\mathrm{out}}}\right)\right] dr,
\label{eq:isomass}
\end{equation}

\noindent where $\rho_{\mathrm{disk}}=\Sigma_{\mathrm{bg}}/H$ is the volumetric disk gas density, given in terms of the gas surface density $\Sigma_{\mathrm{bg}}$ and the disk scale height $H=c_{s}/\Omega$, where $\Omega$ is the Keplerian orbital frequency. The local disk sound speed is $c_{s}=\sqrt{k_bT_{\mathrm{disk}}/ \mu}$. The gas surface density is calculated at the point in the disk's evolution when photoevaporative winds have caused the inner disk to decouple from the outer disk \citep{Lee2022}. The gas of the inner disk then drains out on a viscous timescale and the gas surface density exponentially decreases with time. From \citealt{Lee2022}, this is modeled as

\begin{equation}
    \Sigma_{bg}(a,\Delta t)=10^2 \left(\frac{a}{1 \mathrm
{au}}\right)^{-11/10}\exp{\left(\frac{-\Delta t}{t_{\mathrm{visc}}}\right)},
\end{equation}

\noindent where $\Delta t$ is the time since the disk decoupled and $t_{\mathrm{visc}}$ is the characteristic viscous draining time, fixed to 0.45 Myrs \citep{Lee2022}. The local disk sound speed is dependent on the local disk temperature, which is defined as

\begin{equation}
T_{\mathrm{disk}} = 550\, \mathrm{K} \left(\frac{a}{1 \mathrm{au}}\right)^{-2/5} \left(\frac{L_{\star}}{1.46 L_{\odot}}\right)^{1/5} ,
\end{equation}

\noindent where $L_{\star}$ is the pre-main sequence luminosity of the host star, which we calculate using the MESA Isochrones \& Stellar Tracks (MIST) as a function of stellar mass and age \citep{Dotter2016, Choi2016,Paxton2011, Paxton2013, Paxton2015}. We choose an age of 1 Myr at which to calculate the luminosity. 

The planet's outer radius $R_{\mathrm{out}}$ in Eq.~\ref{eq:isomass} is evaluated as a numerical factor $f_R$ times the minimum of the Bondi or Hill radii, $R_{\mathrm{out}} = f_R \mathrm{min}(R_{H}, R_{B})$. The Hill radius 

\begin{equation}
    R_H = a\left(\frac{M_{\mathrm{core}}}{M_{\star}}\right)^{(1/3)},
    \label{eq:Hill}
\end{equation}

\noindent defines the boundary at which the planet dominates the local gravitational potential. In Eq.~\ref{eq:Hill}, $a$ is the planet's semimajor axis and $M_{\star}$ is the mass of the host star. The Bondi radius is

\begin{equation}
R_B = \frac{GM_{\mathrm{core}}}{c_s^2},
\end{equation}

\noindent is a function of core mass and the local disk temperature.

For a planetary system to be consistent with the primordial radius valley model, the envelope mass fraction ($X_{\mathrm{env}}= M_{\mathrm{iso}}/M_{\mathrm{core}}$) of the enveloped planet must be greater than the envelope mass fraction of the rocky planet after formation (i.e. $X_{\mathrm{rocky}} < X_{\mathrm{env}}$). Our framework calculates the minimum mass of the enveloped planet by solving for the core mass that equates $X_{\mathrm{rocky}}$ and $X_{\mathrm{env}}$. 

Note that the utility of using multi-transiting rocky-enveloped planet pairs to evaluate the primordial radius valley model is evident in Eq.~\ref{eq:isomass}, which depends on the disk volumetric density $\rho_{\mathrm{disk}}$. Because the ages of known multi-transiting systems far exceeds the relevant timescales for gas-accretion (i.e. $\sim 1$ Myr), values of the disk's volumetric density $\rho_{\mathrm{disk}}$ are not directly observable in individual planetary systems. The comparison of $X_{\mathrm{rocky}}$ and $X_{\mathrm{env}}$ in the same planetary system therefore negates the need for a-priori knowledge of $\rho_{\mathrm{disk}}$ and allows us to test the consistency of the primordial radius valley mechanism with multi-transiting systems.

\section{Model Evaluation} \label{sec:code}
\begin{figure*}
\centering
\includegraphics[width=0.95\hsize]{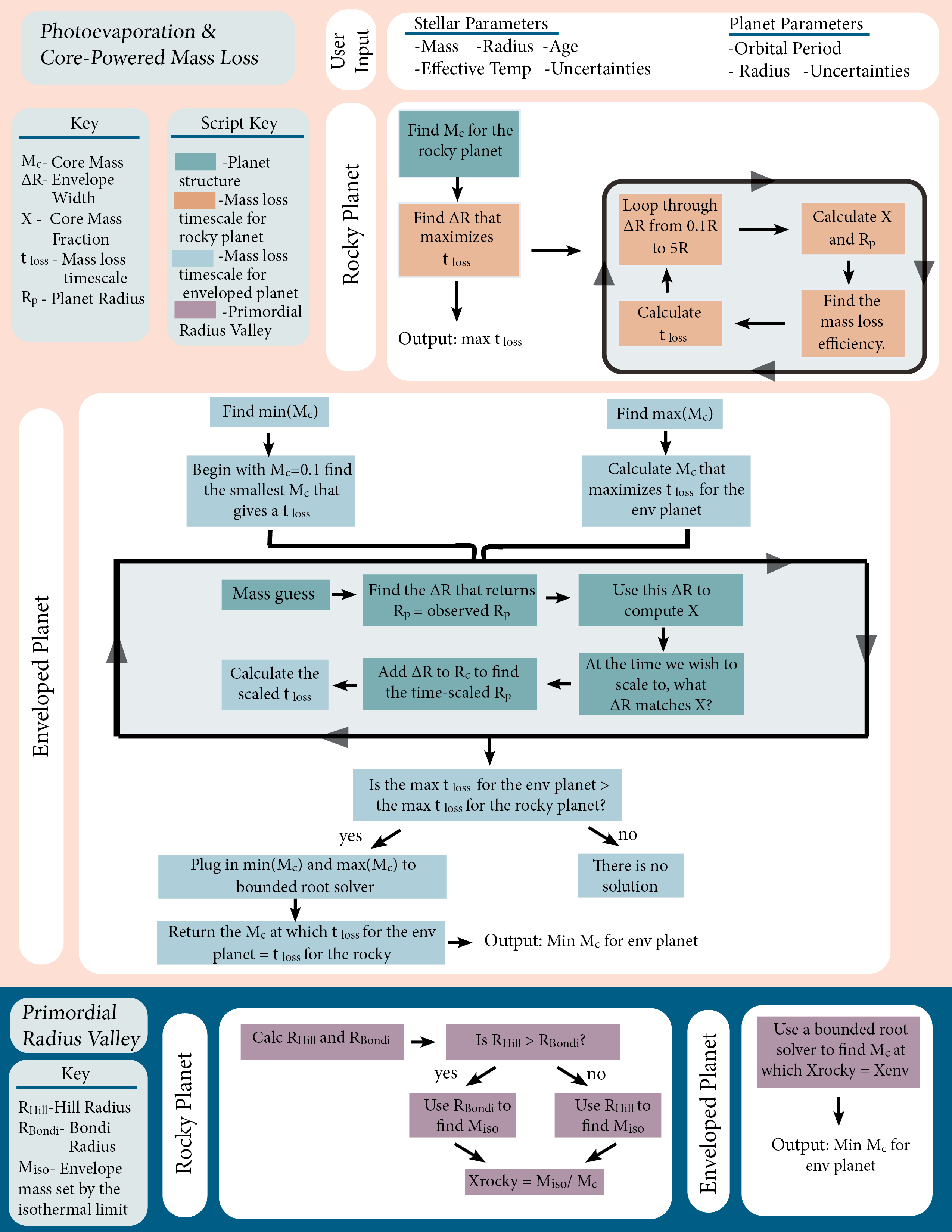}
\caption{\texttt{PEPPER}'s code structure for radius valley emergence models of XUV-driven photoevaporation, core-powered mass loss (top panel), and a primordial radius valley (bottom panel). The code, available on \href{https://github.com/mvanwyngarden/PEPPER}{GitHub}, is broken into four modules, which are color-coded in the key in the top-left of the graphic. The flowchart illustrates the processes performed on both the rocky and enveloped planets within a multi-transiting system.}
\label{fig:flowchart}
\end{figure*}

Here we provide an overview of how we evaluate the consistency of each radius valley emergence model in an individual multi-transiting system. The code to perform these calculations, named \texttt{PEPPER} (Primordial, corE-Powered, or PhotoEvaporation Radius valley), is provided on \href{https://github.com/mvanwyngarden/PEPPER}{GitHub} and archived at doi:10.5281/zenodo.12785928. The description below serves as a guide for the code's usage. The steps are also summarized in the flowchart in Figure~\ref{fig:flowchart}.

\subsection{User Inputs} \label{sec:inputs}
All three models require the user to specify a minimum of eight required planetary and stellar parameters. These include the planetary radii and orbital periods for at least one planet pair that spans the radius valley. Additional sets of radius and period parameters may be specified in multi-transiting systems that contain more than two planets. The user must also specify four stellar parameters: stellar mass, radius, effective temperature, and (approximate) age. Uncertainties on any of the input parameters can be optionally specified by the user and are discussed in Section~\ref{sect:error}.

\subsection{Code Structure: XUV-driven Photoevaporation} \label{sec:PEcode}
Here we provide an outline of the code structure in the context of the photoevaporation model. In forthcoming Sections~\ref{sect:cpml_model} and \ref{sect:prv_model}, we will note differences in the code's structure for the core-powered mass loss and primordial radius valley models, respectively.

For a planetary system to be consistent with photoevaporation, the enveloped planet must have a mass loss timescale greater than the maximum mass loss timescale of the rocky planet. Our goal is therefore to solve for the critical core mass of the enveloped planet for which its mass loss timescale is equal to the maximum mass loss timescale for the rocky planet. The corresponding value of $M_{\mathrm{core,env}}$ is the minimum mass of the enveloped planet needed for the system to be consistent with photoevaporation.

We begin by calculating the maximum mass loss timescale for the rocky planet. Figure 1 of \citealt{OwenWu} demonstrates that an envelope width, $\Delta R$, exists for every planet that maximizes the mass loss timescale. This inflection point, which is not captured by Equation \ref{eq:tmlPE} alone, arises as planets with envelope widths below a certain threshold will have short mass loss timescales that rapidly strip the planet of its envelope. Conversely, above a certain threshold, the radius of the planet can become dominated by the radius of the envelope. In this regime, the envelope is not as tightly bound to the core, resulting in shorter evaporation timescales. 

In order to capture this inflection point, our model loops through $\Delta R$ ranging from 0.1 to 5 times the core radius to find the envelope width that maximizes the rocky planet's mass loss timescale. For each value of $\Delta R$, we calculate the envelope mass fraction, $X_{\mathrm{env}}$, of the planet. The core mass of the planet is given in Section~\ref{sec: planetstructure} and the mass of the envelope is calculated using the same prescription as \citetalias{OwenEstrada},

\begin{equation}
 M_{\mathrm{env}}=4\pi R_{RCB}^3 \rho_{RCB} \left(\nabla_{ab}\frac{G M_c}{c_s^2 R_{RCB}}\right)^{1/\gamma -1} I_2(R_c/R_{RCB}, \gamma).
\end{equation}

\noindent Here $\rho_{RCB}$ is the density of the RCB, $\nabla_{ab} = \gamma -1/ \gamma$ is the adiabatic gradient, $M_c$ is the planetary core mass, $c_s$ is the isothermal sound speed, and $\gamma$ is the adiabatic index set at 5/3. $I_n$ is the dimensionless integral

\begin{equation}
    I_n(R_c/R_{RCB}, \gamma) = \int_{R_c/R_{RCB}}^{1} x^n(x^{-1} -1)^{1/\gamma-1} dx.
\end{equation}

As in \citetalias{OwenEstrada}, we use an opacity law $\kappa = \kappa_0 P^{\alpha}T^{\beta}$ to give the density at the radiative-convective boundary as 

\begin{equation}
    \rho_{RCB} \propto \left(\frac{\mu}{k_b}\right) \left[\left(\frac{I_2}{I_1}\right) \frac{64\pi T_{eq}^{3-\alpha-\beta} R_{RCB} \tau_{KH}}{3 \kappa_0 M_p X}\right]^{1/(1+\alpha)}
\end{equation}

\noindent where $\alpha=0.68$, $\beta=0.45$, and $\kappa_0=4.79\times 10^{-8}$ when pressure and temperature are expressed in cgs units. The envelope mass fraction, the radius of the planet $R_p=R_c+\Delta R$, and core mass are then used to calculate the mass loss efficiency. As we loop through values of $\Delta R$, we evaluate Eq.\ref{eq:tmlPE} to solve for $\Delta R$ that maximizes the mass loss timescale of the rocky planet.

Once the maximized mass loss timescale for the rocky planet is known, we employ a bounded root solver, \texttt{scipy.optimizie.brentq}, to find the enveloped planet's mass at which the difference between the mass loss timescale for the enveloped planet and the maximized mass loss timescale for the rocky planet is zero. We initially set the lower bound on the root solver, $\mathrm{min}(M_c)$, to 0.1 $M_{\oplus}$, but if a mass loss timescale is unable to be calculated at this initial guess then $M_c$ is increased in steps of 0.1 $M_{\oplus}$ until a solution can be found. The maximum mass bound, $\mathrm{max}(M_c)$, is chosen to be the core mass that maximizes the mass loss timescale for the enveloped planet.  We calculate the maximized mass loss timescale for the enveloped planet using a similar procedure to that of the rocky planet. We begin with a $\mathrm{max}(M_c)$ guess and find the $\Delta R$ value required to return the observed enveloped planet radius. $\Delta R$ is used to calculate the planet's envelope mass fraction. Following the phase of severe atmospheric mass loss, the planet's envelope mass fraction will remain nearly constant to the planet's current age. The envelope mass fraction is therefore used to calculate the enveloped planet's radius at a rescaled time in the past to account for thermal contraction (see Section~\ref{sec: thermalevol}). It is this time-scaled radius value that is used to determine the photoevaporative mass loss timescale. We loop through maximum mass guesses until the mass, $\mathrm{max}(M_c)$, that maximizes the mass loss timescale is found. If the max $t_{\mathrm{loss}}$ for the enveloped planet is greater than that of the rocky planet then the root solver is used to locate a mass in the range $[\mathrm{min}(M_c),\mathrm{max}(M_c)]$ that equates the timescales of the rocky and enveloped planets. 

If available, comparing an enveloped planet's measured mass to our model-predicted minimum mass can be used to confirm whether the system's architecture is consistent with photoevaporation.

\begin{deluxetable*}{c c c c c c c c c c c c c}[t!]
\tablecaption{Multi-transiting systems spanning the radius valley and without mass measurements.
\label{tab:systemsnomass}}
\tablehead{
\colhead{System}& \colhead{Planet}&\colhead{$P_{\mathrm{rocky}}$}& \colhead{$R_{p,\mathrm{rocky}}$}&   \colhead{$P_{\mathrm{env}}$}&\colhead{$R_{p,\mathrm{env}}$}&
\colhead{$M_{\star}$}& \colhead{$R_{\star}$} & \colhead{$T_{\mathrm{eff}}$} &\colhead{Min Mass} &\colhead{Min $M_p$} &\colhead{Min $M_p$}\\
\colhead{Name}&\colhead{Pair}&\colhead{(days)}& \colhead{($R_{\oplus}$)} &\colhead{(days)} &\colhead{($R_{\oplus}$)}& \colhead{($M_{\odot}$)} &\colhead{($R_{\odot}$)} & \colhead{(K)} &\colhead{PE} &\colhead{CPML} &\colhead{PRV}}
\startdata
K2-166& c,b & 3.80 & 1.50 & 8.53 & 2.00 & 1.15 & 1.48 & 6084 & $2.45_{-1.77}^{+3.55}$ &  $4.50_{-3.37}^{+5.92}$ & $3.64_{-2.56}^{+4.91}$ \\
K2-16 & d,b & 2.72 & 1.03 &  7.62 & 2.02 & 0.67 &      0.66 & 4341 & $0.62_{-0.46}^{+0.81}$ & $1.48_{-1.14}^{+1.92}$ & $1.00_{-0.83}^{+1.23}$ \\
K2-16& d, c & 2.72 & 1.03 & 19.08 & 2.54 & 0.67 &       0.66 & 4341 & $0.45_{-0.30}^{+0.71}$ & $1.37_{-0.91}^{+2.11}$ & $0.83_{-0.71}^{+1.01}$ \\
EPIC 206024342& b,d &  4.51 & 1.36  & 14.65 &     2.01 & 0.88 &  0.85 & 5559 & $1.41_{-1.04}^{+1.87}$ &  $3.00_{-2.30}^{+3.88}$ & $2.26_{-1.71}^{+2.96}$ \\
EPIC 206024342& c, d & 0.91 & 1.08 &  14.65 &     2.01 & 0.88 & 0.85 &  5559 &  $0.23_{-0.19}^{+0.27}$ & $0.82_{-0.66}^{+1.06}$ & $0.88_{-0.79}^{+1.00}$\\
\enddata
\tablecomments{For conciseness, only a subset of five rows are depicted here to illustrate the table's contents. The entirety of this table is provided in the arXiv source code and will ultimately be available as a machine-readable table in the journal.}
\end{deluxetable*}

\subsection{Code Structure: Core-Powered Mass Loss} \label{sect:cpml_model}
For core-powered mass loss, the overall structure of our code is identical to photoevaporation as described in Section~\ref{sec:PEcode}. The only differences are the mass loss timescale expression (see Section~\ref{section:CPML}) and the time at which we compare the planets' mass loss timescales. Unlike in our photoevaporation model, this timescale is fixed to 1 Gyr in our core-powered mass loss model.

\subsection{Code Structure: Primordial Radius Valley} \label{sect:prv_model}
Our calculations that pertain to the primordial radius valley model do not follow identically to our calculations of thermally-driven mass loss (i.e. photoevaporation and core-powered mass loss). The distinct code structure is summarized at the bottom of Figure~\ref{fig:flowchart}. 

In this model, we use the planetary core mass given in Section \ref{sec: planetstructure} and the orbital period to identify the minimum between the Hill or Bondi radii of the rocky planet and set this radius to $R_{\mathrm{out}}$ to evaluate its maximally-cooled isothermal envelope mass using Eq.~\ref{eq:isomass}. The rocky planet's envelope mass fraction is then taken to be the ratio of its maximally-cooled isothermal mass limit and its core mass ($X_{\mathrm{rocky}} = M_{\mathrm{iso}} / M_{\mathrm{core}}$). To test the consistency of the primordial radius valley model with a multi-transiting system, we again employ the root solver \texttt{scipy.optimizie.brentq} to solve for the minimum core mass of the enveloped planet that equates its envelope mass fraction $X_{\mathrm{env}}$ to $X_{\mathrm{rocky}}$.

\begin{figure*}
\includegraphics[width=\textwidth]{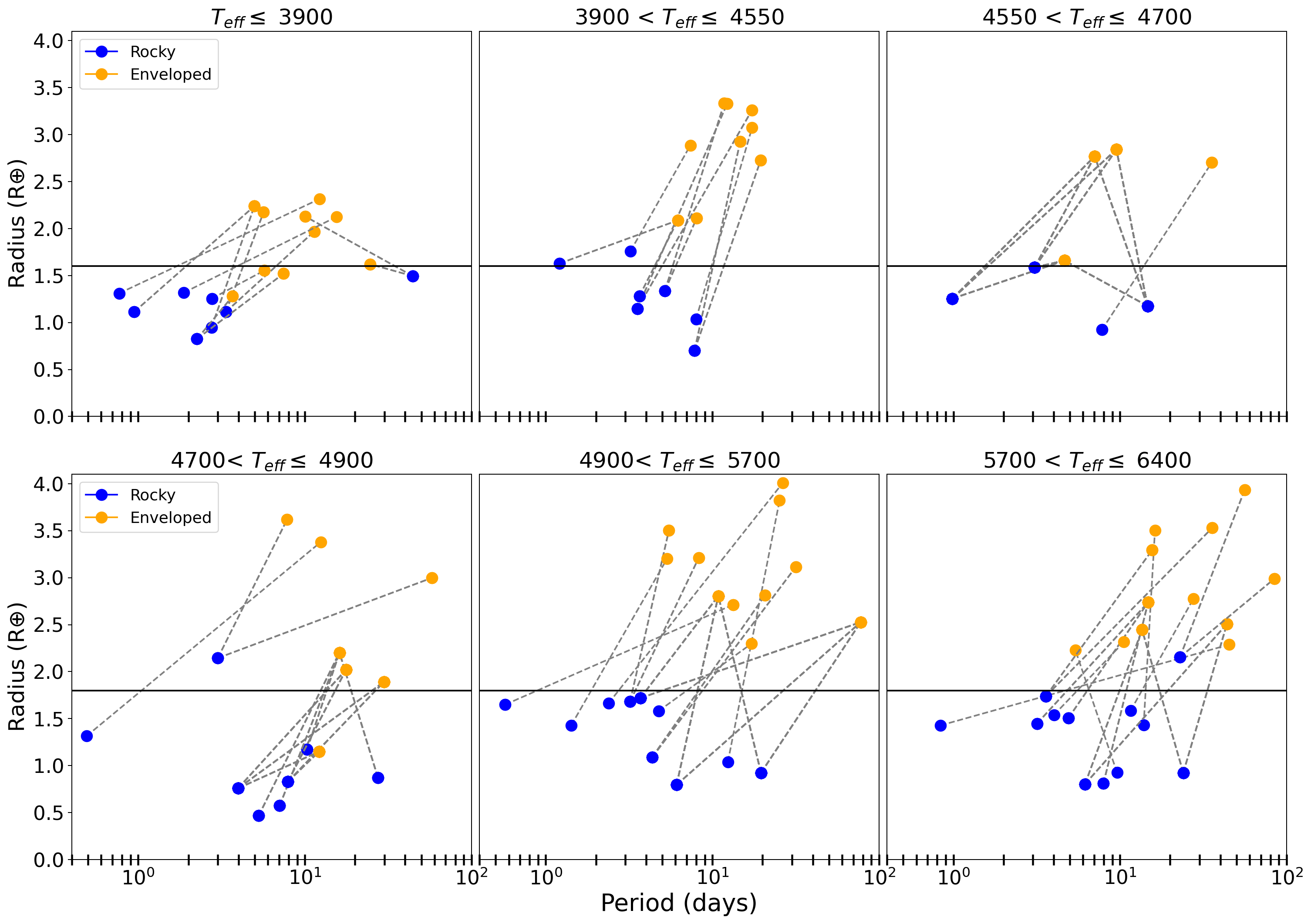}[t]
\caption{The planetary pairs with measured masses used in this study binned by stellar effective temperature. Dashed lines connect rocky-enveloped pairs within the same planetary system. The location of the radius valley is plotted as a black line at 1.6 $R_{\oplus}$ for $T_{\mathrm{eff}} < 4700$ K and at 1.8 $R_{\oplus}$ for $T_{\mathrm{eff}} > 4700$ K.} 
\label{fig:planets}
\end{figure*}

\subsection{Inclusion of Input Parameter Errors} \label{sect:error}
When evaluating the minimum enveloped planet's mass under any of the models within \texttt{PEPPER}, the calculation can be performed taking into account the uncertainties on the required input parameters. The user may choose to specify the $1\sigma$ uncertainties on any or all of the required parameters, along with the desired number of realizations $N$. When errors are provided and $N>1$, a Gaussian distribution is adopted from which $N$ values of the input parameters are sampled. In this case, the evaluation of each model will yield a distribution of $N$ enveloped planet minimum masses. The distribution of model-predicted minimum masses may then be compared to the enveloped planet's mass measurement posterior to probabilistically assess the consistency of the system's parameters with the physical model in question.

\section{Application to Confirmed Systems and Results} \label{sec:methods}
Here we apply our models to the set of confirmed multi-transiting systems. In systems for which the enveloped planet's mass has been measured, we highlight occurrences of inconsistencies between the modeled minimum masses and the mass measurements.

\subsection{Planetary Sample Selection}
To identify multi-transiting systems that apply to our models, we first queried the NASA Exoplanet Archive to select confirmed multi-planet systems \citep{2013PASP..125..989A, https://doi.org/10.26133/nea12}. We restricted our selection to close-in multi-transiting systems with multiple planets whose orbital periods are $<100$ days, radii are $<4\, R_{\oplus}$, and with radius detection significances $\geq 3\sigma$. We then cut systems that do not have all the required stellar and planetary parameters available, with the exception of stellar ages, which are undetermined for the majority of host stars in our sample (see Section \ref{sec:inputs}). We adopt the refined stellar parameters from \cite{Berger23} including estimates of stellar age. For the 18 systems missing from the sample in \cite{Berger23}, we followed their methods to make the necessary corrections to those outstanding stellar parameters and assign a default age of $5 \pm 2$ Gyr to be consistent with the field population.


After the preliminary cuts to our planet sample, we select only the multi-planet systems whose planets span the radius valley. We identified 210 planets with measured masses, which we are able to classify as either `rocky' or `enveloped' by comparing their masses and radii to mass-radius relations of MgSiO$_3$ and Fe from \cite{Zeng2013}. We perform this classification following the classification scheme of \citep{Cherubim_2023} whereby rocky planets are flagged as those whose measured masses exceed the mass of a pure MgSiO$_3$ body (i.e. CMF=0\%), at the planet's radius. To account for planets' mass and radius uncertainties, we sampled each planet's joint mass-radius posterior $10^3$ times and flagged rocky planets as those that satisfy the aforementioned condition in $\geq 60$\% of samples. Planets whose measured masses were less than the mass of a pure MgSiO$_3$ body in $\geq 60$\% of samples were flagged as being gas-enveloped. All remaining planets were marked as ambiguous and removed from the planetary sample. We note that our sample included six planets whose masses and radii produced ultra-high CMF values that are inconsistent with expectations from mantle stripping by giant impacts \citep{Marcus2010}. We deemed these high-density planets as having unphysical bulk compositions and eliminated them from our sample. Our planet sample with measured masses is depicted as a function of radius, period, and stellar effective temperature in Figure \ref{fig:planets}.

For the remaining 991 planets in our sample that do not have mass measurements reported in the literature, we cannot rely on the classification scheme described above. We aim to classify these planets as either rocky or enveloped based on the side of the radius valley that they are most likely to occupy. However, the true location of the radius valley is expected to vary along many axes in a high dimensional space that includes radius, instellation, stellar mass, and age \citep{Gupta2019,Berger2020,Rogers2021,Sandoval2021}. Because empirical constraints on the radius valley's dependence along each of these axes are severely limited by measurement uncertainties, we refrain from defining the precise location of the radius valley in such a high dimensional space. Instead, we adopt a simplified view of the radius valley's location that is a piecewise function of planet radius and stellar effective temperature only. For systems around stars with $T_{\mathrm{eff}}\geq 4700$ K, we define a fixed location of the radius valley at $1.8\, R_\oplus$ \citep{Fulton2017}. For systems around lower mass stars with $T_{\mathrm{eff}} < 4700$ K, we shift the radius valley's location to $1.6\, R_\oplus$ (\citealt{CloutierMenou}). We use these definitions to label all planets in our sample that lack mass measurements as either rocky or enveloped. 

Our final sample contains 221 multi-planet systems with a total of 389 rocky-enveloped planet pairs. We proceed with calculating the minimum enveloped planet mass for each planet pair under models of XUV-driven photoevaporation, core-powered mass loss, and a primordial radius valley.

\subsection{Results} \label{sec:results}
Evaluation of each rocky-enveloped pair produces a distribution of minimum mass estimates for each radius valley emergence mechanism. The results of our minimum enveloped planet mass calculations for all three models are reported in Table~\ref{tab:systemsnomass}. Systems with large minimum enveloped planet masses but lacking measured masses could be potential targets for observational follow-up to confirm inconsistency with a given radius valley emergence mechanism.

\subsubsection{Failed Models} \label{sect:failed}
Not all combinations of input stellar and planetary parameters are guaranteed to return a minimum planet mass that is consistent with the observed masses and radii of both planets in the system and that satisfies the model's criteria outlined in Section~\ref{sec: data}. These ``failed models'' return NaN values for the minimum mass of the enveloped planet and occur when no mass can be found for which the planet could have accreted and maintained an envelope.  We denote a failed model as one in which $>50\%$ of model realizations returned a NaN value. In Table \ref{tab:systemsnomass}, failed models are denoted with a minimum enveloped planet mass of $20\, M_{\oplus}$. This mass was arbitrarily chosen to represent that an unphysically large mass is required for that model to be consistent with the planet's observed mass and radius.

In our total sample, 44 out of 389 planetary pairs failed to find a solution for photoevaporation. In 34 of these pairs, the rocky planet is at a larger orbital period than the enveloped planet such that it receives a lower incident XUV flux than the system's enveloped planet. Equation 4 of \citetalias{OwenEstrada} demonstrates that the mass required of the enveloped planet under photoevaporation scales as the ratio of the semimajor axis of the rocky planet to the semimajor axis of the enveloped planet. When the rocky planet is exterior to the enveloped planet, a larger mass is required, making the system more difficult to reconcile with the photoevaporation model. It is possible that if the rocky planet is less massive than what our model assumes (i.e. a mass corresponding to a CMF=33\% described in Section \ref{sec: planetstructure}) then the enveloped planet may no longer require an unphysically large mass to be consistent with photoevaporation and the number of failed systems could decrease. In 18 out of the 44 failed planet pairs, the rocky planet has a radius that is $\geq 75\%$ that of the enveloped planet. As the difference in radii decreases between the two planets, it becomes less likely that a solution exists wherein the enveloped planet is able to retain its H/He envelope after the rocky planet has lost one. We also find that 18 out of 389 planet pairs fail to find a solution under the core-powered mass loss model. Of these, 16 pairs feature the rocky planet at a larger period such that it exhibits a lower equilibrium temperature, making it less susceptible to hydrodynamic escape than the hotter enveloped planet. There are 13 out of 18 of these planet pairs where the rocky planet has a radius $\geq 75\%$ of the enveloped planet. Only 8 out of 389 pairs failed to find a solution under the primordial radius valley model. Six of these planet pairs feature the rocky planet at a larger period. Again, if the rocky planet is less massive than what our model assumes, the number of failed systems under core-powered mass loss and the primordial radius valley model could decrease. 

In total, seven planetary pairs failed to find a single solution for all of the proposed mechanisms: K2-3 d/c, Kepler-80 d/e, K2-84 c/b,  Kepler-220 e/c, Kepler-296 e/b, Kepler-326 c/b, and Kepler-342 c/b. The first two of these systems have existing mass measurements and are further described in Section~\ref{sec:k2_3}. The remaining systems are potential targets for observational follow-up as their system architectures may be explained by alternative radius valley emergence mechanisms not considered in this study. 

\begin{figure}[t]
\includegraphics[width=\hsize]{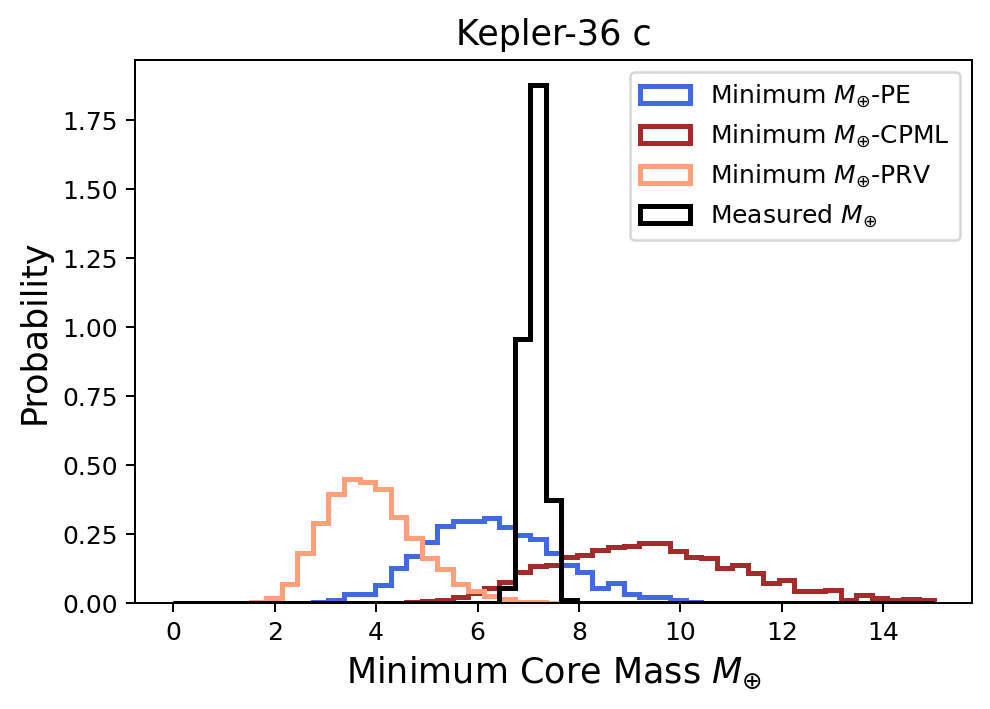}
\caption{Model-predicted minimum core masses of Kepler-36 c in order for the Kepler-36 system to be consistent with radius valley emergence mechanisms of XUV-driven photoevaporation (blue), core-powered mass loss (red), and a primordial radius valley (green). The minimum core mass distributions are shown alongside the planet's mass posterior of $7.13 \pm 0.18\, M_{\oplus}$ from \cite{Vissapragada}.} 
\label{fig:kep36}
\end{figure}

\subsubsection{Comparison with Measured Masses}
75 planet pairs in 36 planetary systems have measured masses for the enveloped planet. We wish to compare each model's predicted minimum mass to the known planet masses to evaluate the probability of each system's consistency with each radius valley emergence model. We do so by drawing $10^6$ samples from each model's minimum mass distribution and from the enveloped planet's mass and associated uncertainty from the literature. We assume that the latter is well-described by a Gaussian distribution. We define the probability that a planet pair is consistent with a given model as the fraction of model-predicted minimum mass draws that are less than the corresponding measured mass draw. The model consistency probabilities for planets with measured masses are reported in Table~\ref{tab:example systems}.

\begin{figure*}[t] 
\centering
\includegraphics[scale=0.72]{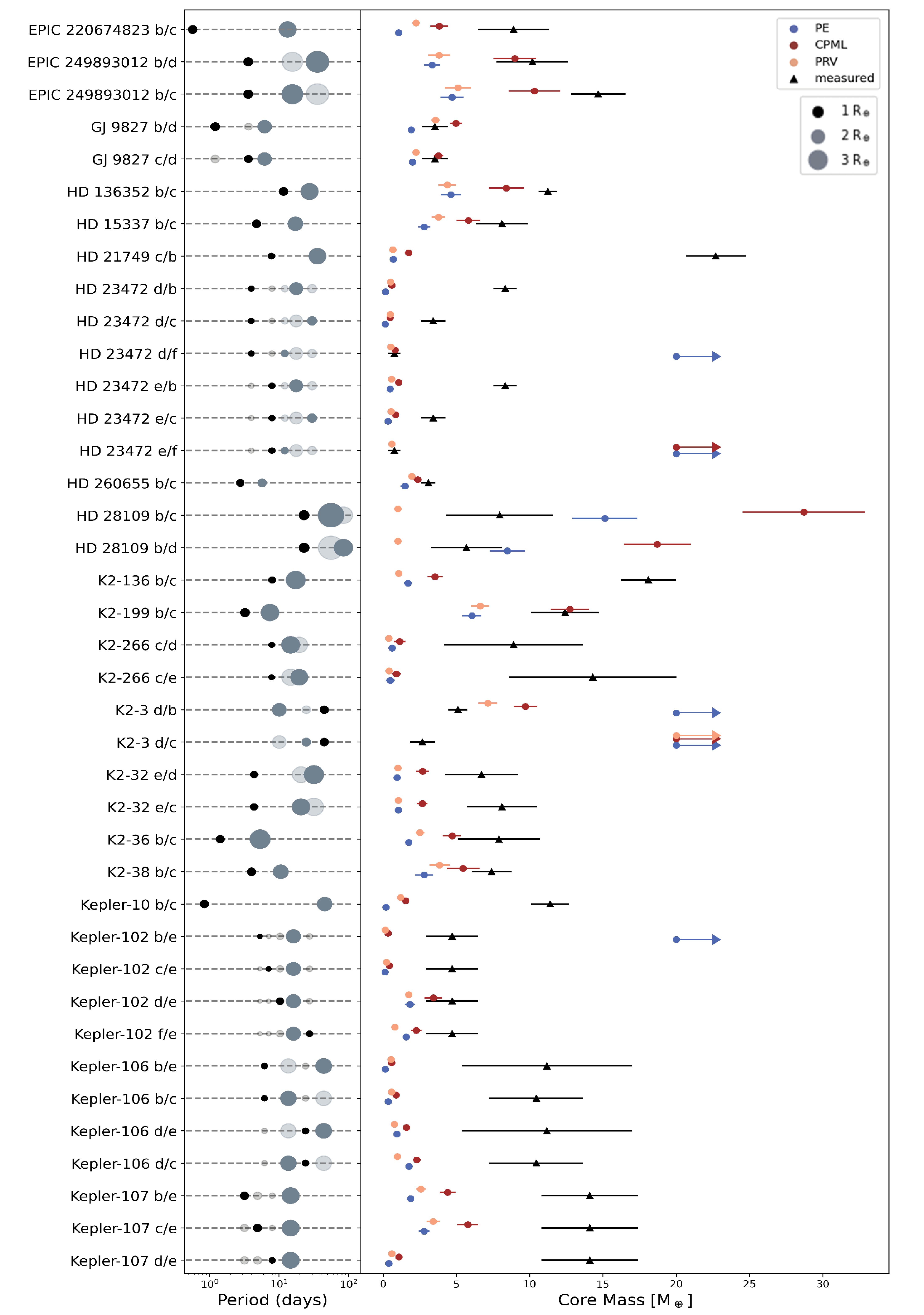}
\caption{The architectures and model-predicted minimum masses for half of the 75 planet pairs in our sample of systems for which the enveloped planet mass is known. The left panel displays the systems' orbital architectures and planet sizes. Mutual inclinations are not reflected. Planets are classified as being either rocky (black markers) or gas-enveloped (grey), based on their observed masses and radii. In a given row, the system's planets that do not belong to the rocky-enveloped pair being considered are translucent. The right panel compares the model-predicted minimum masses of the enveloped planets in each pair for three radius valley emergence models: XUV-driven photoevaporation (PE; blue circles), core-powered mass loss (CPML; red circles), and a primordial radius valley (PRV; green circles). The model-predicted minimum masses are compared to the enveloped planet's measured mass (black triangles). All point estimates include their respective $1\sigma$ errorbars.}

\label{fig:resultsfull}
\end{figure*}

\begin{figure*} 
\centering
\includegraphics[width=0.95\hsize]{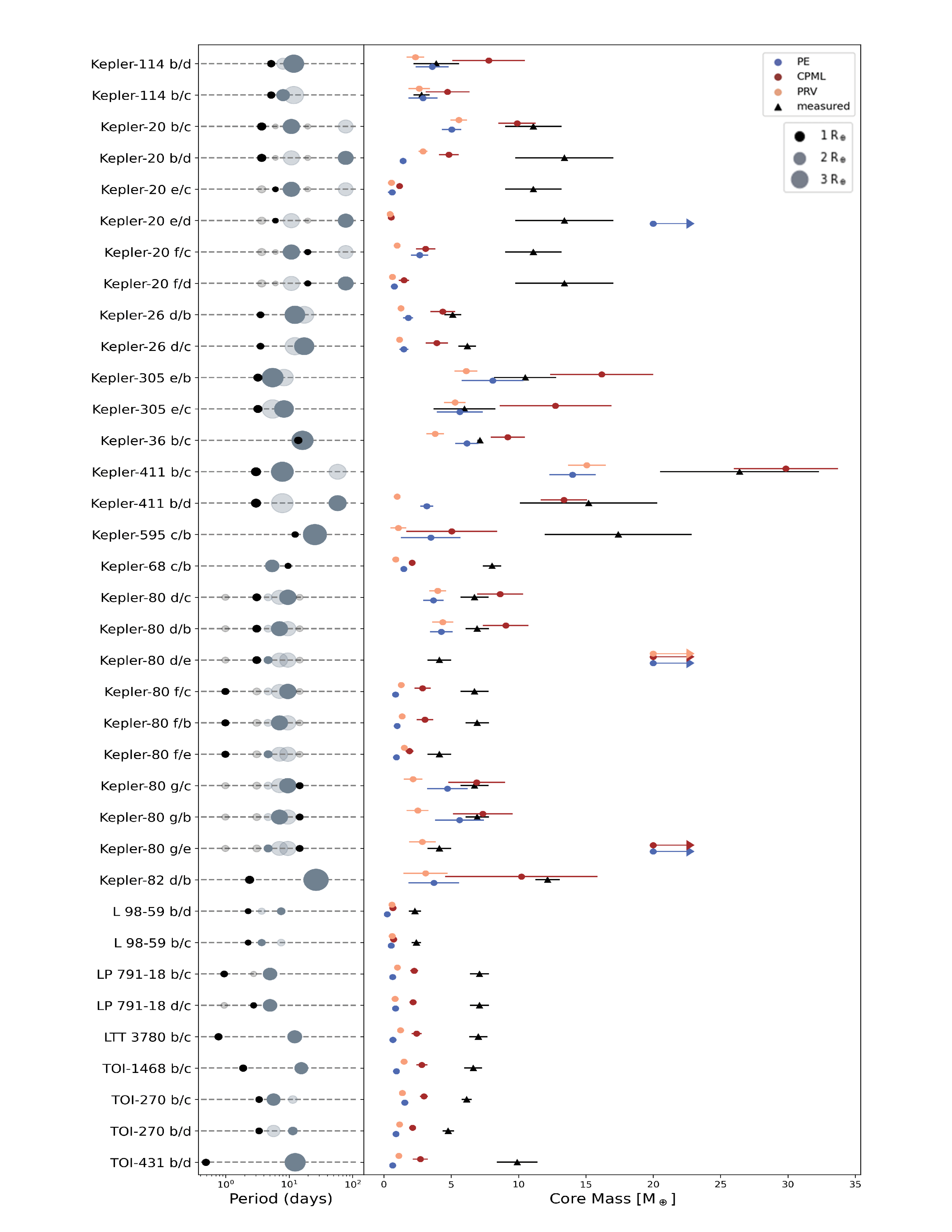}
\caption{Same Figure~\ref{fig:resultsfull} for the second half of our planet sample.}
\label{fig:resultsfull2}
\end{figure*}

We show an example of our model comparison process for the well-studied Kepler-36 system in Figure \ref{fig:kep36}. Kepler-36 is a two-planet system consisting of both a rocky and an enveloped planet with large differences in radii but similar orbital distances. As both planets would have experienced a similar XUV instellation history, this system is a particularly interesting test case for XUV-driven photoevaporation. The rocky planet, Kepler-36 b, has a radius of $1.43 \pm 0.09 R_{\oplus}$ and an orbital period of 13.87 days, while the enveloped planet c has a radius of $3.50 \pm 0.16 R_{\oplus}$ and an orbital period of 16.22 days. We generate the distribution of minimum masses of the system's enveloped planet for each model using 3000 samples of its stellar and planetary parameters from \cite{Vissapragada}. We calculate the probability that the system is consistent with photoevaporation to be 74.6\%. This is in agreement with previous hydrodynamic simulations of this system, which also concluded that XUV-driven photoevaporation could plausibly account for the system's unique configuration \citep{LopezFortney, OwenMorton}. Conversely, we find a much lower probability of consistency with the core-powered mass loss scenario of 11.2\%, suggesting that not all forms of atmospheric escape can explain the Kepler-36 system's architecture. We also find a 99.8\% probability of consistency with the primordial radius valley mechanism.

Figures~\ref{fig:resultsfull} and~\ref{fig:resultsfull2} summarize the results of our modeling effort applied to all 75 planet pairs that have measured masses for the enveloped planet. The figures are structurally identical and are split for visual clarity. These figures compare point estimates of the model-predicted minimum masses to the measured masses and allow for an illustrative assessment of model consistencies for individual planet pairs. We include failed models (see Section~\ref{sect:failed}) in Figures~\ref{fig:resultsfull} and~\ref{fig:resultsfull2}. While these systems failed to find a solution using our model's framework in over half of the realizations, the reason for this is that an unphysically large mass is required for that model to be consistent with the observed planetary parameters. These systems are difficult to reconcile with model prediction and so we illustrate the unphysically large mass requirements by placing the model's minimum mass at $20\, M_{\oplus}$ in Figures~\ref{fig:resultsfull} and ~\ref{fig:resultsfull2} and replacing the minimum mass uncertainties with an arrow.

\section{Discussion} \label{sec:discussion}
The multi-transiting systems that we consider span a range of system architectures and host stellar properties. We find that the majority of the planet pairs in our sample can be readily explained by any of the three mechanisms. These systems are not useful for investigating the prevalence of any physical mechanism because model consistency does not imply that the model is responsible for producing the observed architecture. However, there are several systems for which consistency with a given mechanism is unclear or largely inconsistent. These are the interesting systems that may allow us to rule out the prevalence of a particular mechanism in some regions of the planetary-stellar parameter space. Here we examine model consistencies over our full sample before examining model consistencies as a function of stellar mass and stellar metallicity. 

We note that systems containing multiple pairs of rocky-enveloped planets will exhibit different probabilities of consistency with a given mechanism. For example, one planet pair may be consistent with photoevaporation while another pair exhibits inconsistency. We deem such planetary systems as inconsistent with the mechanism under consideration. We therefore adopt the minimum probability across a system's planet pairs when we report the system's probability of consistency with each mechanism.

\subsection{Searching for Trends in Model Consistency} \label{sect:trends}
\begin{figure*}[t]
\includegraphics[width=\textwidth]{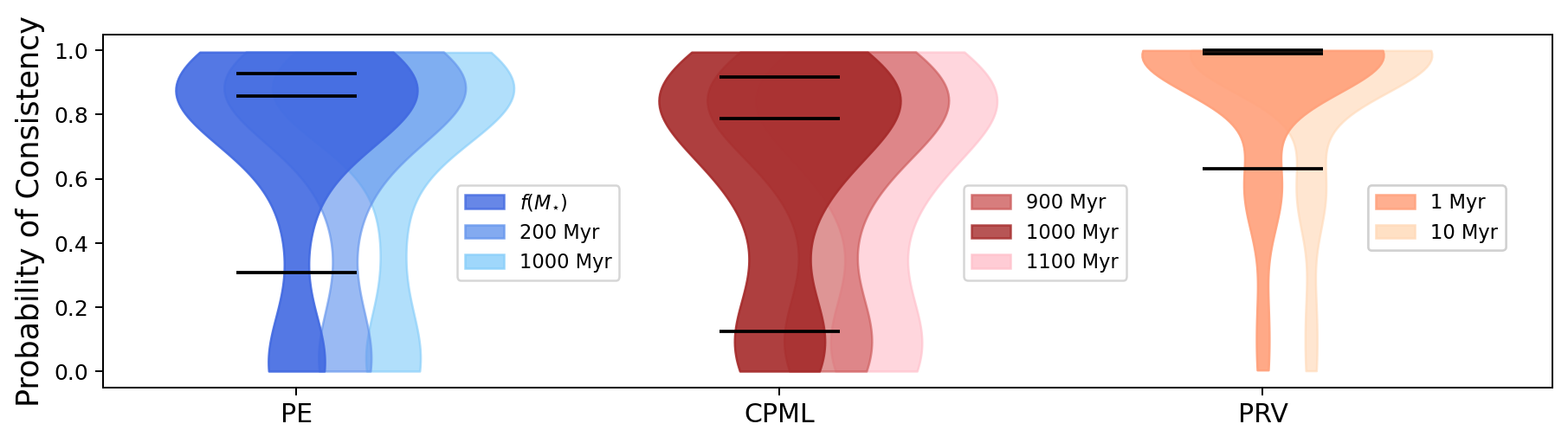}
\caption{Distributions of the probability of planetary system consistency with a given mechanism for our full sample of 36 multi-transiting systems. The three physical mechanisms are abbreviated by PE (photoevaporation), CPML (core-powered mass loss), and PRV (primordial radius valley). Each distribution is smoothed by a kernel density estimation and the horizontal markers highlight the 16th, 50th, and 84th percentiles, respectively. Translucent violins depict the results of testing additional timescales than those described in Section \ref{sec: thermalevol}. The fiducial timescale for PE, labeled $f(M_\star)$, is a function of stellar mass that ranges from 100-1000 Myrs.} 
\label{fig:violin_all}
\end{figure*}

The distribution of consistency probabilities for our full sample of 36 systems is shown in Figure \ref{fig:violin_all}. The violin plots depict kernel density estimates of the underlying distribution of model consistency probabilities for each of the photoevaporation, core-powered mass loss, and primordial radius valley models. We find that for all three mechanisms, the bulk of the consistency probability distributions are $\gtrsim 70$\% with median consistency probabilities of 86\%, 79\%, and 99\% for photoevaporation, core-powered mass loss, and a primordial radius valley, respectively. This suggests that the majority of systems are consistent with all three radius valley emergence mechanisms. We note that the core-powered mass loss scenario has the lowest fraction of consistent systems while the primordial radius valley has the highest fraction. We also test additional options for the time at which the rocky and enveloped planets are compared in each mechanism. In Section \ref{sec: thermalevol} we select a timescale for each mechanism, shown by the opaque violins in Figure \ref{fig:violin_all}, to ensure that the two planets are being compared at the same stage in their thermal evolution. Other choices of timescale are tested, depicted by the translucent violins, to confirm that the choice of timescale does not discernibly affect our results.

\begin{figure*}[t] 
\includegraphics[width=\textwidth]{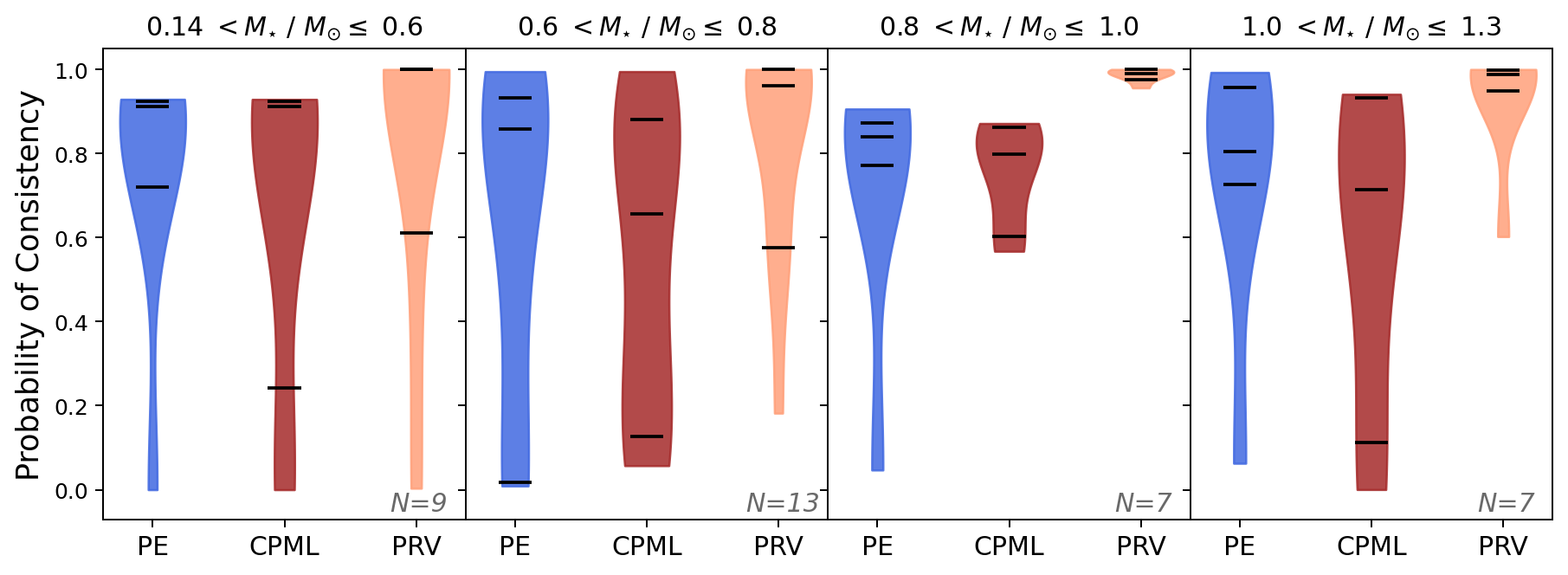}
\caption{Same as Figure~\ref{fig:violin_all} after binning our planet sample into four stellar mass bins. The number of systems $N$ per bin is annotated in each panel.} 
\label{fig:violin_mass}
\end{figure*}

Figure \ref{fig:violin_mass} depicts the probability of model consistency for each mechanism after binning our planetary systems by host stellar mass. This investigation is warranted because the efficacy of each mechanism is expected to vary with stellar mass \citep{Rogers2021,Lee2022}, which is supported by observations showing that the slope of the radius valley in the period-radius parameter space varies between planets around FGK stars versus lower mass K and M dwarfs \citep{Martinez2019,CloutierMenou,Petigura2022}.  
The first stellar mass bin of $[0.14,0.6]\, M_\odot$ roughly corresponds to M dwarfs, while the three subsequent bins increase in stellar mass by 0.2 $M_\odot$. This is similar to the binning adopted by \cite{FultonPetigura2018}. 

Across our full range of stellar masses, the primordial radius valley model has the highest percentage of consistency, with 16th percentiles $>50$\% and median consistency probabilities of $>90$\%. The small number of outliers implies that the primordial radius valley model can plausibly explain the existence of nearly all known multi-transiting systems of rocky-enveloped planet pairs. Conversely, the models of thermally-driven atmospheric escape appear to show some random variation with stellar masses although any trends are not statistically significant given the small number of multi-transiting systems in each bin. We do note that the median probabilities for photoevaporation and core-powered mass loss appear to increase and decrease in tandem from bin to bin. This is unsurprising given that the underlying physical processes are very similar between these mass loss models, with the only major differences arising from their exact scalings with stellar and planetary parameters. Overall, we find no significant evidence that supports that one mechanism dominates over others.

\begin{figure*}[t] 
\includegraphics[width=\textwidth]{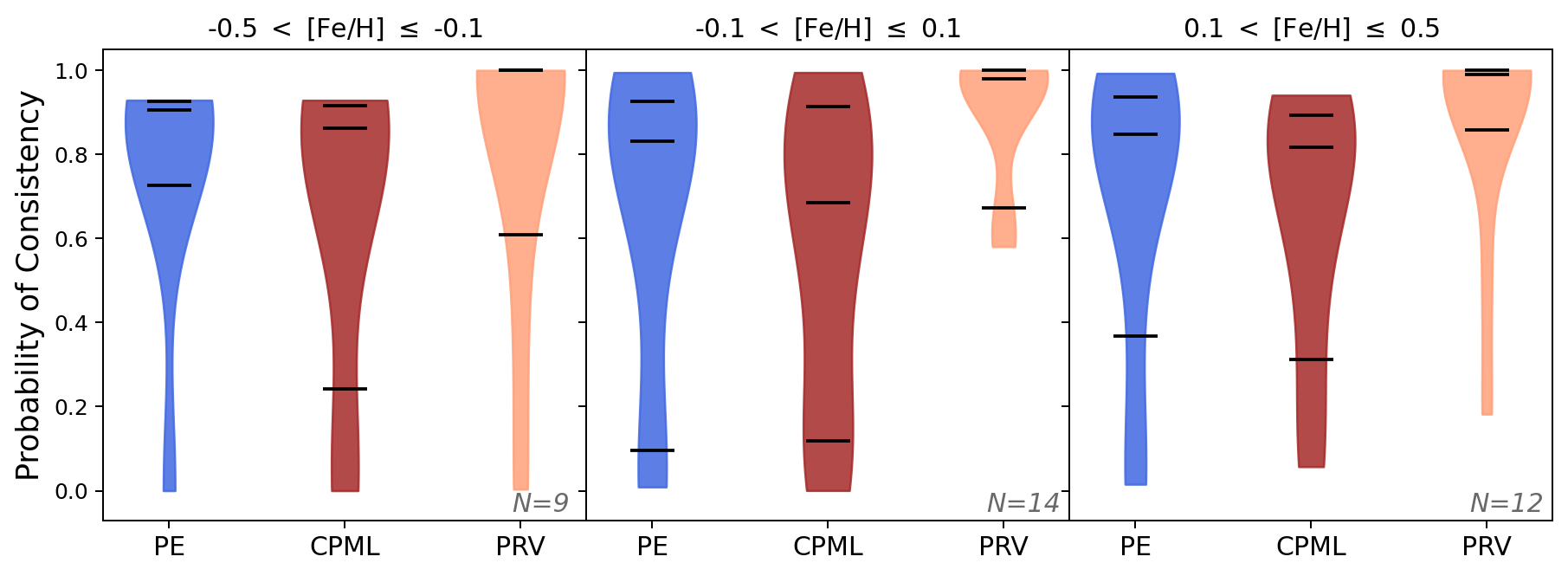}
\caption{Same as Figure \ref{fig:violin_all} after binning our planet sample into three stellar metallicity bins. The number of systems $N$ per bin is annotated in each panel.} 
\label{fig:violin_met}
\end{figure*}

Lastly, we examine the probability of model consistency as a function of stellar metallicity in Figure \ref{fig:violin_met}. We continue to see that the primordial radius valley model boasts the highest probabilities of consistency, although there is no apparent trend with stellar metallicity. The median probabilities for photoevaporation and core-powered mass continue to vary in unison across metallicity bins, although no broad trend in model consistency as a function of stellar metallicity is apparent. We conclude that more multi-transiting systems with rocky-enveloped planet pairs with precisely measured masses will be required before multi-transiting systems can serve as a viable diagnostic of radius valley emergence models.

\subsection{Systems with Universal Inconsistency} \label{sec:k2_3}
In this section, we analyze planetary systems that are inconsistent with all three proposed emergence mechanisms. Our sample of 36 systems contains only two such planetary systems whose mass measurements result in probabilities of consistency less than 40\% for all three mechanisms: K2-3 and Kepler-80.

K2-3 is a three-planet system around an M0 dwarf with two inner enveloped planets and an outer (assumed) rocky planet \citep{Crossfield2015}. The mass of the outer planet only has an upper limit of $1.6\, M_{\oplus}$, but has a small radius of $1.49 \pm .05$ that places it below the radius valley, suggestive of a rocky bulk composition \citep{Bonomo2023}. The system's architecture, with its enveloped planets on smaller orbits than the rocky planet, is atypical compared to the bulk of multi-transiting systems with rocky-enveloped pairs. Such configurations can be challenging to reconcile with radius valley emergence models. In the K2-3 system, both pairs of rocky-enveloped planets are inconsistent with both photoevaporation and core-powered mass loss in all model realizations and show only a 0.3\% consistency with the primordial radius valley model. \cite{Diamond-Lowe} provide a compelling explanation for the inconsistency of the K2-3 system with XUV-driven photoevaporation, which may extend to reconciling the system with the other two mechanisms. \cite{Diamond-Lowe} analyzed the planets' masses, radii, and their XUV environment using HST/COS spectra and data from XMM-Newton. They argued that H/He envelopes of the two inner planets b and c are unstable to hydrodynamic escape driven by the star's XUV luminosity. The interpretation is that K2-3 b and c are not H/He-enveloped and instead are water-rich worlds that formed beyond the water ice line before undergoing inward planetary migration. This scenario has been supported by planet formation theory around low mass stars \citep{Burn2021} and is discussed more generally in Section~\ref{sect:water}.

Kepler-80 is a six-planet system with three rocky planets and three enveloped planets orbiting a K5 star \citep{MacDonald2016,Shallue2018}. In order of increasing orbital period, the rocky planets occupy positions one, two, and six, positioning the outermost rocky planet beyond the orbits of the enveloped planets (c.f. Figure~\ref{fig:resultsfull2}). Though multiple planet pairs show inconsistency with at least one mechanism, it is the planet pair d-e, occupying orbital positions two and three, that is inconsistent with all three mechanisms. This arises because the planets are closely separated ($P_d=3.07$ days, $P_e=4.64$ days) with consistent radii ($R_{p,d}=1.59 \pm 0.11, R_{p,e}=1.66\pm 0.11, R_{\oplus}$), but exhibit a large mass discrepancy ($M_{p,d}=6.75 \pm 0.60\, M_{\oplus}, M_{p,e}=4.13\pm0.88\, M_{\oplus}$), that make their respective mass loss timescales and $M_{\mathrm{iso}}$ values diverge and become inconsistent with all three mechanisms. This could indicate that the assumed atmospheric structure of the planet is incorrect. As shown in \citep{Sandwiched2024}, 
the timescale for the formation of this chain of small planets compared to the evolutionary timescale for the evolution of the snow line could lead to compositional degeneracies, something we may be witnessing in the Kepler-80 system. In addition to the d-e pair, the g-e pair only shows consistency with photoevaporation in $<3$\% of model realizations. 
Of the remaining pairs, four are consistent with core-powered mass loss in $<35$\% of realizations. Outside of the d-e pair, the system shows consistency with the primordial radius valley model in $>$ 70\% of realizations. 

\subsection{Systems with Partial Consistency}
There are two planetary systems that are consistent with the primordial radius valley model but not photoevaporation or core-powered mass loss. The first, HD 23472, is a 5-planet system orbiting a K4 dwarf that features two super-Mercuries and three super-Earths and is the only known system to harbor more than one super-Mercury \citep{Barros2022}. The complete architecture of the system is consistent with the primordial radius valley model, but planet pair e-f failed to find solutions for photoevaporation and core-powered mass loss, and planet pair d-f has $< 40\%$ probability of consistency with models of thermally driven mass loss. Like the K2-3 system, \cite{Barros2022} note that these inconsistencies may be alleviated if the systems' three sub-Neptunes are actually water worlds.

The second system is HD 28109, a 3-planet system with one rocky and two enveloped planets orbiting an F6 dwarf \citep{Dransfield2022}. Both rocky-enveloped planet pairs are consistent with the primordial radius valley model only. However, the system's planets have long periods of 23, 56, and 84 days, resulting in long photoevaporative mass loss timescales. We calculated the absolute XUV-driven mass loss timescale using a dimensional version of Eq.~\ref{eq:tmlPE} and assuming that the host star had an XUV luminosity of $5\times 10^{-4} L_{\mathrm{bol}}$ up to its saturation time, where $L_{\mathrm{bol}}$ is the star's bolometric luminosity. We found that the rocky planet at 23 days has a mass loss timescale of 391 Myrs, while the two enveloped planets have mass loss timescales of 415 and 840 Myrs. These timescales are likely a few times longer than the host star's saturation timescale of $<100$ Myrs \citep{McDonald2019}, implying that photoevaporation is unlikely able to strip any of these planets' atmospheres. The same reasoning likely follows for other thermally-driven mass loss processes like core-powered mass loss. While it is possible that XUV-driven atmospheric mass loss can continue past the star's saturation timescale \citep{KingWheatley}, atmospheric escape around sub-Neptunes has almost exclusively been seen around young stars with ages less than 1 Gyr \citep{Bennett2023}. It is worth noting that \cite{Orell2024} argues there is no preference to observe this atmospheric escape in young systems. Even in the rare cases where atmospheric escape has been observed on a Gyr timescale \citep{Zhang2023}, it is impossible to attribute the mass loss to photoevaporation or core-powered mass loss.

Two systems appear inconsistent with photoevaporation only (Kepler-102 and Kepler-20). There are an additional five systems that appear inconsistent with just core-powered mass loss (GJ 9827, Kepler-114, Kepler-305, Kepler-36, Kepler-411). These systems highlight the interplay between two different mechanisms for thermally-driven atmospheric escape, which may be driven by the details of the planet's atmospheric structures. In particular, the extent of their Bondi radii relative to the XUV penetration depth, which dictates the interplay between photoevaporation and core-powered mass loss \citep{OwenSchlichting}.

\subsection{Comparison to \texttt{EvapMass}}
In this work, we build upon the assumptions and photoevaporation model presented in \citetalias{OwenEstrada} and the accompanying code \texttt{EvapMass}. \texttt{PEPPER} differs in a few key ways. The primary difference is that our model is capable of testing two additional proposed radius valley emergence mechanisms: core-powered mass loss and a primordial radius valley model. Additionally, the inaugural version of \texttt{EvapMass} featured a bug in the treatment of the mass loss efficiency parameter in its photoevaporative mass loss timescale. This bug did not significantly impact the results presented in \citetalias{OwenEstrada} and has been corrected in recent updates to \texttt{EvapMass}. Lastly, the inaugural version of \texttt{EvapMass} adopted an efficiency parameter $\eta$ that scaled with the planet's escape velocity as $\propto v_{\mathrm{esc}}^{-2}$. When applied to the variety of known multi-transiting systems, our preliminary tests yielded unphysical values for $\eta>1$, which is why \texttt{PEPPER} has options for alternative $\eta$ prescriptions outlined in Section~\ref{section:PE}. Recent updates to \texttt{EvapMass} also include similar flexibility for the user to specify their desired prescription for mass loss efficiency.

\subsection{Caveats to our Model}
Our model seeks to test three radius valley emergence mechanisms under a number of assumptions regarding the formation and composition of radius valley planets. These model assumptions are not unique to our work as they have been commonly assumed or acknowledged in the formulation of each of the aforementioned models. In the following subsections, we outline these caveats and their potential impact on our inferences presented in Section~\ref{sect:trends}.

\subsubsection{Location of the radius valley}
The radius valley is expected to be highly dimensional, with its location varying with radius, instellation, stellar mass, age, and other potential factors \citep{Gupta2019,Berger2020,Rogers2021,Sandoval2021}. However, the exact location of the radius valley along each of these axes is uncertain. This makes the task of determining on which side of the valley a transiting planets falls on when its mass is unknown, inexact. The exact classification is particularly ambiguous for planets close to the expected location of the valley. Therefore, when labeling planets as rocky or enveloped, we chose to consider only a projection of the radius valley in the planetary radius and effective temperature plane, where the location of the valley has been empirically constrained \citep{FultonPetigura2018,CloutierMenou}. While this choice matters for the classification of planets lacking mass measurements, these are not the planets that can be used to infer model consistencies. As such, our exact choice of the radius valley's location does not affect our emergence model inferences discussed in Section~\ref{sect:trends}.

\subsubsection{Degeneracy of Planetary Compositions} \label{sect:water}
For planets with measured masses, distinguishing between rocky versus gas-enveloped compositions was based on the planets' masses and radii compared to internal structure models \citep{Zeng2013}. However, the exact bulk composition of low-density planets is degenerate when based on mass and radius measurements alone. One major degeneracy is water worlds. Some of the planets classified as gas-enveloped in our work may instead be water worlds rather than rocky cores surrounded by thick H/He envelopes. This is especially true for the hottest low-density planets in our sample whose H/He envelopes would likely be lost to thermally-driven escape. While these planets' compositions cannot be observationally established without atmospheric characterization, in this study we have assumed that all low-density planets are H/He-enveloped terrestrial cores. This assumption follows from the evidence for this interpretation based on the observed location and structure of the radius valley \citep{OwenWu,Gupta2019,Wu2019,Bean2021}. But we note that water worlds, or alternative planetary compositions other than rocky and enveloped, may exist in our sample and would add a new degree of complexity to our model assessment that we do not consider in this paper. 

Contamination of our planet sample by water worlds may be particularly important in planetary systems around M dwarfs. Several lines of recent empirical evidence suggest that the larger planet peak in the radius valley around M dwarfs, unlike around FGK stars, may be populated by water worlds rather than H/He-enveloped terrestrials. Arguments that favor the water world interpretation include the slope of the M dwarf radius valley suggesting a primordial origin \citep{CloutierMenou}, detailed studies of individual M dwarf planetary systems that are inconsistent with models of thermally driven escape \citep{Diamond-Lowe,Piaulet2023}, and characterizing the population of planets in or spanning the M dwarf radius valley \citep{LuquePalle,Cherubim_2023,Bonfanti2023}. In this case, the M dwarf radius valley must arise as rocky planets form within the snow line and the lower-density water worlds form beyond it and migrate inward \citep{Venturini2020,Burn2021}. 

\subsubsection{Planetary migration}
Our evaluation of each proposed radius valley emergence model is dependent on planets' orbital distances, or some proxy thereof. Because planet migration histories are difficult to ascertain in dynamically cold, non-resonant systems, we do not include the effects of planetary migration in our models. Instead, we assume that the planets in our sample have not undergone migration since the epochs relevant to the formation of the radius valley. 

This simplification is insensitive to any form of disk-driven migration in the photoevaporation and core-powered mass loss scenarios, both of which tend to operate after the dispersal of the protoplanetary disk. However, this assumption may not be valid in the primordial radius valley scenario because the formation of the valley is contemporaneous with the epochs of disk-driven migration. Assessing the impact of gas accretion during migration would require careful modeling of both processes, which is beyond the scope of this paper. However, we note that this assumption may still be valid given the demonstration that the primordial radius valley model, in the absence of planetary migration, does show good agreement with the observed radius valley around FGK stars \citep{Lee2022}. 

We also note that by assuming that planets in our sample do not migrate after disk dispersal, we are ignoring any form of non-disk-driven migration such as planet-planet scattering or Lidov-Kozai interactions \citep{Weidenschilling1996,Wu2003}. These migration mechanisms are dynamically hot and could shift planets' orbital distances after disk dispersal, consequently rendering our model evaluations inaccurate. However, we argue that this is likely a minor effect given the prevalence of compact systems of multiple small planets that do not show evidence of high eccentricity migration \citep{VanEylen2019,He2020,Albrecht2022,Sagear2023}.

\section{Conclusion} \label{sec:conclusion}
We presented a test of three different radius valley emergence mechanisms using systems of multi-transiting planets that span the radius valley. Our work expands upon the work of \citetalias{OwenEstrada} and their \texttt{EvapMass} model of XUV-driven photoevaporation. Our model, called \texttt{PEPPER}, is made available on \href{https://github.com/mvanwyngarden/PEPPER}{GitHub} and can be used to assess the consistency of multi-transiting system architectures with XUV-driven photoevaporation, core-powered mass loss, and the primordial radius valley models. For the thermally-driven atmospheric mass loss mechanisms of photoevaporation and core-powered mass loss, our model calculates the minimum mass needed for an enveloped planet in order to retain its atmosphere given that the system's rocky planet has lost its envelope. The primordial radius valley calculation returns the minimum mass required by the enveloped planet to accrete a more massive H/He envelope mass fraction than the system's rocky planet.

We applied our model to 221 systems containing 389 rocky-enveloped planet pairs. Within this sample, 75 planet pairs have (assumed) gas-enveloped planets with mass measurements that we compared to our model predictions to evaluate the consistency of planetary systems with each emergence mechanism. We found that the majority of planetary systems are consistent with all three mechanisms, though the primordial radius valley model has the highest median probability of consistency across the full sample. Additionally, there are no observable trends between model consistency with either stellar mass or metallicity. We do find a handful of notable cases in which a planetary system is consistent with only one or two of the proposed mechanisms, but not all. 

For planetary systems lacking mass measurements, the minimum mass estimates returned by our models can be used to identify candidates for observational follow-up. Additionally, our models may be applied to any forthcoming multi-transiting system discoveries that contain a rocky-enveloped planet pair that spans that radius valley. As transit surveys continue to identify new multi-planet systems, this framework may be used to probe their formation and evolution and may ultimately serve as a unique diagnostic of radius valley emergence mechanisms if many more multi-transiting systems have their masses characterized.

\section{Acknowledgments}
This project began as a part of the Smithsonian Astrophysical Observatory's (SAO) Research Experience for Undergraduates (REU) program. The SAO REU program is funded in part by the National Science Foundation and Department of Defense ASSURE programs under NSF Grant no.\ AST- 2050813, and by the Smithsonian Institution. 

The authors thank Collin Cherubim and James Owen for fruitful discussions of hydrodynamic escape processes and James Owen for assistance with interpreting the \texttt{EvapMass} code.

\bibliography{mybib.bib}

\newpage
\begin{longrotatetable}
\begin{deluxetable*}{c c c c c c c c c c c c c c c }
\tablecaption{Multi-transiting systems spanning the radius valley with mass measurements used to compute a probability of consistency with each radius valley emergence mechanism.
\label{tab:example systems}}
\tablehead{
\colhead{System}& \colhead{Planet} &\colhead{$P_{\mathrm{rocky}}$}&\colhead{$R_{p,\mathrm{rocky}}$}& \colhead{$P_{\mathrm{env}}$}& \colhead{$R_{p, \mathrm{env}}$}& \colhead{$M_{p,\mathrm{env}}$} &
\colhead{$M_{\star}$}& \colhead{$T_{\mathrm{eff}}$} &\colhead{Min $M_p$} &\colhead{Min $M_p$} &\colhead{Min $M_p$} &\colhead{PE} &\colhead{CPML} &\colhead{PRV}\\
\colhead{Name}&\colhead{Pair}& \colhead{(days)} &\colhead{($R_{\oplus}$)}& \colhead{(days)} &\colhead{($R_{\oplus}$)}& \colhead{ ($M_{\oplus})$} & \colhead{($M_{\odot}$)} & \colhead{(K)} &\colhead{PE} &\colhead{CPML} &\colhead{PRV} &\colhead{Prob} &\colhead{Prob} &\colhead{Prob}
}
\startdata
EPIC 220674823 & b, c & 0.57 & 1.65 &    13.34 &  2.71 &  $8.90\pm{2.4}$ & 0.98 & 5618 & $0.99_{-0.76}^{+1.21}$ & $3.63_{-3.01}^{+4.31}$ &     $2.24_{1.96}^{2.58}$ &  0.81 &  0.80 & 1.00 \\
EPIC 249893012 & b, c & 3.60 & 1.74 &   15.62 &  3.29 & $14.67_{-1.89}^{+1.84}$ &  1.13 &    5826 & $4.71_{-3.67}^{+6.06}$ &      $10.31_{-8.12}^{+13.30}$ & $5.10_{-3.94}^{+6.65}$ & 1.00 & 0.88 & 1.00 \\
EPIC 249893012 & b, d &  3.60 &  1.74  &  35.75 & 3.53 & $10.18_{-2.42}^{+2.46}$ &  1.13 &     5826 &  $3.34_{-2.57}^{+4.15}$ &   $8.97_{-6.90}^{+11.14}$ & $3.82_{2.81}^{4.98}$ &   0.99 &      0.64 &     0.99 \\
GJ 9827& b, d & 1.21 &  1.63  &  6.20 & 2.09 &   $3.53_{-0.88}^{+0.87}$ & 0.60  & 4063 &   $1.89_{-1.69}^{+2.14}$ & $4.91_{-4.42}^{+5.47}$ & $3.57_{-3.18}^{+4.02}$ &  0.88 &  0.08 &  0.47 \\
GJ 9827& c, d & 3.65 & 1.28 & 6.20 & 2.09 & $3.53_{-0.88}^{+0.87}$ &  0.60 &   4063 &$1.97_{1.76}^{2.22}$ &  $3.72_{3.31}^{4.16}$ &  $2.25_{2.02}^{2.50}$ & 0.87 &  0.38 & 0.92\\
\enddata
\tablecomments{For conciseness, only a subset of five rows are depicted here to illustrate the table's contents. The entirety of this table is provided in the arXiv source code and will ultimately be available as a machine-readable table in the journal.}
\end{deluxetable*}
\end{longrotatetable}

\end{document}